\documentclass[useAMS,usenatbib,twocolumn]{mnras}
\usepackage{graphicx,url,amssymb,amsmath,color,units,wasysym,epsfig,epstopdf,enumerate,tabularx, subfigure}

\usepackage{newtxtext,newtxmath}
\usepackage[T1]{fontenc}
\usepackage{ae,aecompl}

\usepackage[dvipsnames]{xcolor}
\usepackage[normalem]{ulem}
\usepackage[bottom]{footmisc}

\renewcommand{\sc}[1]{\textsc{#1}}

\newcommand{\new}[1]{{#1}}

\begin{document}

\title[Searching for Eccentricity]{Searching for Eccentricity: Signatures of Dynamical Formation in the First Gravitational-Wave Transient Catalogue of LIGO and Virgo}

\author[Romero-Shaw, Lasky \& Thrane]{
Isobel M. Romero-Shaw$^{1,2}$\thanks{isobel.romero-shaw@monash.edu}, Paul D. Lasky$^{1,2}$ and Eric Thrane$^{1,2}$
\\
$^{1}$Monash Centre for Astrophysics, School of Physics and Astronomy, Monash University, VIC 3800, Australia\\
$^{2}$OzGrav: The ARC Centre of Excellence for Gravitational-wave Discovery, Clayton, VIC 3800, Australia
}

\maketitle

\begin{abstract}
Binary black holes are thought to form primarily via two channels: isolated evolution and dynamical formation.
The component masses, spins, and eccentricity of a binary black hole system provide clues to its formation history. 
We focus on eccentricity, which can be a signature of dynamical formation.
Employing the spin-aligned eccentric waveform model \sc{SEOBNRE}, we perform Bayesian inference to measure the eccentricity of binary black hole merger events in the first Gravitational-Wave Transient Catalogue of LIGO and Virgo.
We find that all of these events are consistent with zero eccentricity.
We set upper limits on eccentricity ranging from 0.02 to \new{0.05} with 90\% confidence at a reference frequency of $\unit[10]{Hz}$.
These upper limits do not significantly constrain the fraction of LIGO-Virgo events formed dynamically in globular clusters, because only $\sim5\%$ are expected to merge with measurable eccentricity. 
However, with the Gravitational-Wave Transient Catalogue set to expand dramatically over the coming months, it may soon be possible to significantly constrain the fraction of mergers taking place in globular clusters using eccentricity measurements.
\end{abstract}


\section{Introduction}
\label{sec:intro}

The first Gravitational-Wave Transient Catalogue (GWTC-1) of Advanced LIGO~\citep{Aasi13} and Virgo~\citep{AdvancedVirgo} records eleven gravitational-wave signals, each of which was produced by the coalescence of compact stellar remnants \citep{GWTC-1}. 
The question of how these binaries formed has become paramount. 
With perhaps ${\cal O}(100)$ events expected following the third observing run of Advanced LIGO and Virgo, we are rapidly accumulating the data required to answer this question.

It is challenging to explain how compact binaries form with separations small enough to merge within the age of the Universe.
Most viable scenarios fall into two categories: \textit{isolated binary evolution} and \textit{dynamical formation}. 
The two categories are distinguishable because the formation history of a binary is imprinted on its component masses, component spins, and eccentricity. These binary parameters can be probed using gravitational waves.
In this paper, we make steps towards identifying the formation channels of binary black hole mergers in GWTC-1 using measurements of eccentricity.

The isolated evolution scenario begins with a binary star system. In order to merge within the age of the Universe, the stars must be extremely close; two $\sim\unit[10]{M_{\odot}}$ compact objects in a quasi-circular orbit must have a separation less than $\sim\unit[0.1]{AU}$ to merge within $\sim\unit[14]{Gyr}$  \citep{Celoria18}. Normal stellar evolution prevents binary compact object formation at such small distances since stars expand and consume nearby companions as they age. A number of processes have been proposed to avoid this problem.
The common envelope hypothesis allows the binary components to co-evolve within the extended gas structure of one expanded star (see, e.g., \citet{Livio88, Bethe98, Ivanova13, Kruckow16}).
The chemically homogeneous pathway bypasses the expansion problem, with both stars remaining relatively compact throughout their entire evolution \citep{deMink10, deMink16}. Ambient gas-driven fallback has also been suggested to harden initially distant binaries \citep{Tagawa18}. 

In the dynamical formation case, the merger progenitors do not encounter each other until they are already compact objects. Binaries assemble through encounters in dense environments, such as young star clusters, globular clusters and galactic nuclei; see, e.g., \citet{Sigurdsson93} and \citet{PortegiesZwart99}, plus recent  works such as \citet{OLeary05, Samsing13, Morscher15, Gondan17, Samsing17, Rodriguez18b, Randall17, Randall18, SamsingDOrazio18, Samsing18, Rodriguez18a, Fragione18, Fragione19b} and \citet{Bouffanais19}. Binaries that form in such environments can interact frequently, and one compact object can swap in and out of many binaries before it merges with another compact object. During an interaction between a binary and a single black hole, gravitational binding energy from the incoming binary tends to be converted into the kinetic energy of whichever object leaves the interaction unbound. This leaves the resultant binary with a smaller separation than the binary that entered the interaction. 

The component spins of a binary can be used to distinguish between formation channels (see, e.g., \citet{Belczynski01, Vitale15,   Rodriguez16, Farr17, Fishbach17a, Bianchi18, Wysocki18}).
An isolated binary is likely to be observed with component spins that align with its orbital angular momentum vector due to the co-evolution of the components \citep{Campanelli06, Stevenson17dlk}.
Dynamically-formed binaries have no spin preference, due to their chaotic interactions, so we expect them to be detected with an isotropic distribution of spin orientations (e.g., \citet{Rodriguez16, TalbotThrane17}).

The mass distribution of a merger population may give some insight into its dominant formation channel (e.g., \citet{Stevenson15bqa, Zevin17, TalbotThrane18}). Pulsational pair instability supernovae restrict an isolated merger's total mass to $\lesssim 80M_{\odot}$ \citep{HegerWoosley02, Fishbach17}, whilst mass segregation and runaway mergers in dense environments lead to an extended tail out to high masses for dynamical mergers \citep{Gerosa17, Rodriguez19, Bouffanais19}.

Perhaps the most compelling evidence for dynamical binary formation, however, is eccentricity. Due to the efficient loss of energy through gravitational-wave emission, long-lived binaries circularise rapidly, so we expect binaries from this channel to have negligible eccentricities when they enter the LIGO-Virgo band at $\sim\unit[10]{Hz}$ \citep{Peters64, Hinder07}. 

Dynamically-formed binaries can have a wide range of eccentricities at $\unit[10]{Hz}$, with some having eccentricities close to unity \citep{Zevin17, Rodriguez18b, Gondan18, Samsing18, Zevin18}.
These systems go from formation to merger much faster than their isolated counterparts --- fast enough to retain significant orbital eccentricity when the gravitational-wave frequency reaches $\unit[10]{Hz}$. 
By studying the spin, mass, and eccentricity distributions of the mergers we detect with gravitational waves, we can build a concordant picture of compact binary formation.
\new{It is possible for Kozai-Lidov resonance \citep{Kozai62, Lidov62} to drive up the eccentricity of binaries within hierarchical field triples  \citep{Silsbee16, Antonini17, Fishbach17a, Rodriguez18jqu, triplespin, Liu19} and quadruples \citep{quadruples, quadruples2}, leading them to merge with eccentricity and spin distributions similar to those expected for dynamical mergers. The fraction of mergers from Kozai-Lidov resonance in the field is highly uncertain, although it is expected to be small unless natal kicks and/or environment metallicities are low \citep{Silsbee16, Antonini17, Rodriguez18jqu, triplespin, Liu19}.}

Whilst compact binaries with negligible eccentricity near merger can form by either channel, a single event with significant eccentricity ($e \gtrsim 0.1$) would provide a strong argument for dynamical formation.  Furthermore, the eccentricity of a binary can indicate which subset of dynamically-formed binaries it belongs to.
Dynamically-formed binaries that are ejected from their host cluster are expected to circularise in the field, eventually reaching $e \sim 10^{-6}$ at $\unit[10]{Hz}$ \citep{Rodriguez18a}.
Compact objects that remain in the dense cluster core may form triples and quadruples, which can experience chaotic resonant interactions \citep{Wen02, Antonini15, Rodriguez18a, Rodriguez18b, Samsing18}.
Binaries that harden during such interactions can merge before their next strong encounter, and have an eccentricity distribution that peaks at $e \sim 10^{-4}$ at $\unit[10]{Hz}$ \citep{Rodriguez18a, Rodriguez18b, Zevin18}. During a close dynamical encounter between two compact objects in a globular cluster, the strong loss of gravitational energy at periapsis can lead to a gravitational-wave capture merger.
 In the simulations of \citet{Samsing17}, \citet{Rodriguez18a} and~\citet{Rodriguez18b}, this kind of binary enters the LIGO-Virgo band with $10^{-3} \lesssim e \lesssim 1$. These simulations suggest that we can expect $\sim5\%$ of dynamically-formed binaries to have $e \geq 0.1$ at $\unit[10]{Hz}$, a prediction that is thought to be relatively robust to the assumptions of the globular cluster model.
 The largest values of eccentricity, $e \sim 1$, are obtained when a binary forms with a gravitational-wave frequency already greater than $\sim$ $\unit[10]{Hz}$.

\citet{Lower18} carried out Bayesian inference on simulated gravitational-wave data using an eccentric waveform template, improving upon the Fisher-matrix-type approach demonstrated by \citet{Gondan17abc}. The former study found that GW150914-like events with eccentricities $\gtrsim0.05$ at $\unit[10]{Hz}$ could be distinguished from quasi-circular events using an Advanced LIGO and Virgo detector network at design sensitivity. However, the \sc{EccentricFD}~\citep{eccentricFD} waveform used to obtain these results models only the inspiral, leaving out merger physics. This  waveform also neglects spin effects. Additionally, this analysis was not applied to real data. \citet{EccentricCWB19} conducted an unmodelled search on real data from Advanced LIGO's first two observing runs.
No candidate events were observed (beyond the binaries previously described using quasicircular templates in GWTC-1).
Moreover, the search in~\cite{EccentricCWB19} is unable to provide a measurement of eccentricity.

In this work, we present the first measurements of eccentricity for binary events detected by Advanced LIGO and Virgo. Using spin-aligned waveforms with inspiral, merger, and ringdown, we construct posterior distributions for eccentricity for ten binary black hole merger events. In order to reduce the computational resources required to perform the computationally intensive analysis, we employ a ``likelihood reweighting technique'' from~\cite{Payne} that enables us to introduce an extra parameter, eccentricity, in post-processing.
We find that all of the events in GWTC-1 are consistent with zero eccentricity.
We obtain event-specific upper limits at 90\% confidence ranging from \new{0.024 to 0.054} at a reference frequency of $\unit[10]{Hz}$.

The remainder of this paper is structured as follows.
We outline our analysis methods, including our Bayesian inference approach and post-processing procedure, in Section \ref{sec:methods}. 
We validate our methodology with an injection study in Section \ref{sec:injection}.
We present our results in Section \ref{sec:results}, and discuss these results in the context of dynamical binary formation in Section \ref{sec:discussion}. 

\section{Method}\label{sec:methods}

\begin{figure}
    \centering
    \includegraphics[scale=0.6]{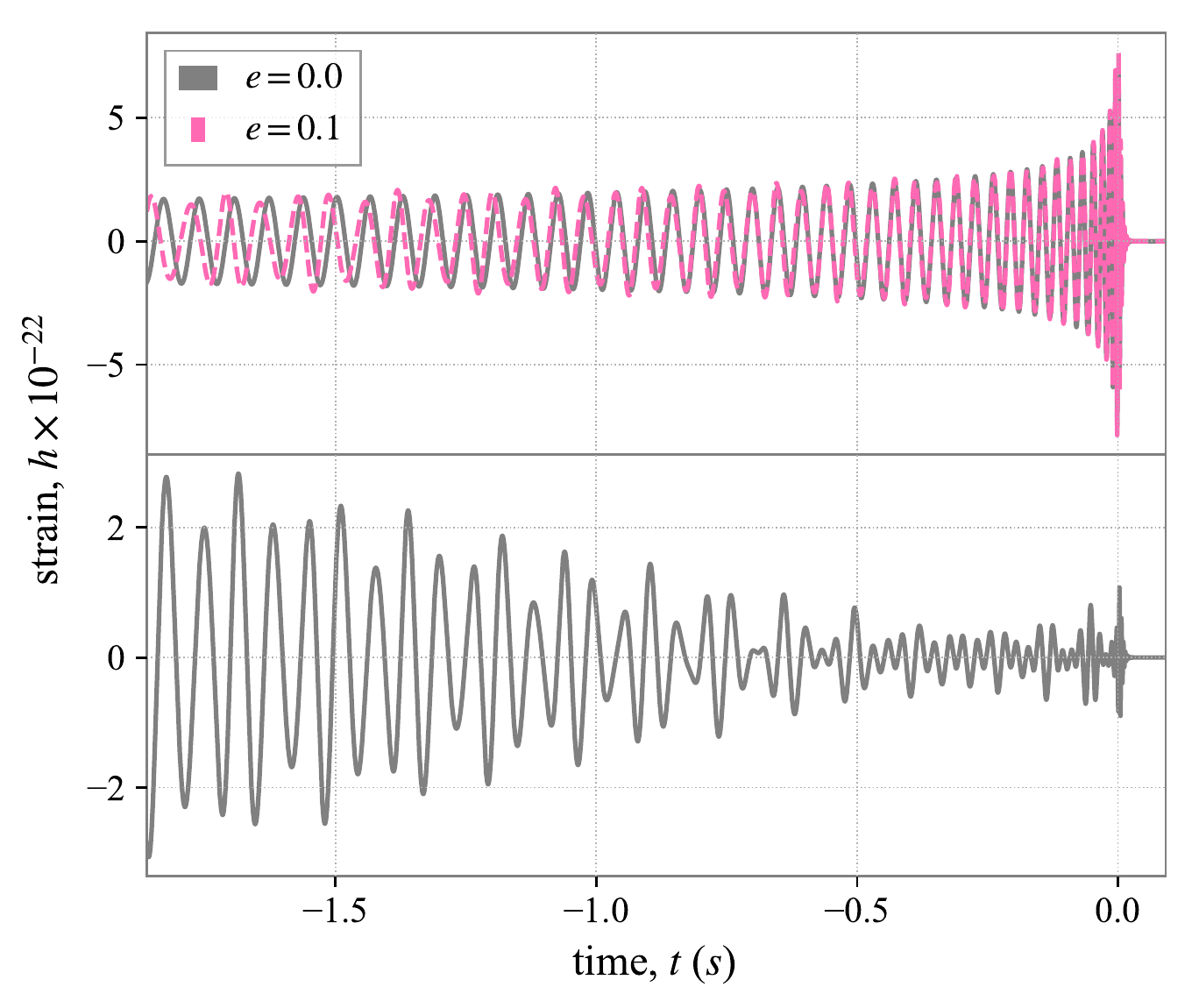}
    \caption{TOP: Gravitational waveforms with eccentricities $e=0.0$ (solid grey) and $e=0.1$ (dashed pink) at $\unit[10]{Hz}$ for a GW150914-like signal. The inclusion of a small eccentricity introduces a slight amplitude and phase modulation, which is most prominent in the early inspiral. BOTTOM: The difference between the quasi-circular and eccentric waveforms as a function of time.}
    \label{fig:waveform_and_residuals}
\end{figure}

Aligned-spin gravitational waveform models usually depend on eleven parameters: four intrinsic (component masses and spins) and seven extrinsic (e.g., luminosity distance and binary inclination angle).
Including eccentricity increases the number of dimensions to twelve. The additional variable is the eccentricity, $e$, at some reference frequency, which we choose to be $\unit[10]{Hz}$.
The gravitational energy released by an eccentric binary at periapsis is greater than that released at apoapsis, so non-zero eccentricity modulates the gravitational-wave signal. The effect of a small binary eccentricity of 0.1 is shown in Figure~\ref{fig:waveform_and_residuals}.

For non-precessing systems, the gravitational waveform depends only trivially on the argument of periapsis because it can be absorbed into the phase of coalescence. 
The situation is more complicated for precessing binaries, but since none of the events in GWTC-1 exhibit clear signs of precession, the effect of precession is likely to be small for published LIGO-Virgo binaries.
At present, there are not publicly available gravitational waveform approximants for eccentric binaries that include precession, although new waveforms are under development; see, for example, \citet{Tiwari19}. 

We use Bayesian inference to measure the parameters describing the binary.
The posterior probability, $p(\theta | d)$, describes the probability that the model with source parameters $\theta$ is responsible for data $d$. The posterior is the product of the likelihood of $d$ occurring if the source model is described by $\theta$, $\mathcal{L}(d | \theta)$, and our prior knowledge of the probability of $\theta$ occurring at all, $\pi(\theta)$. 
Normalising by the model evidence, $\mathcal{Z} = \int d\theta \mathcal{L}(d | \theta)\pi(\theta)$, we can write the posterior probability as
\begin{equation}
\label{eq:bayes}
    p(\theta | d) = \frac{\mathcal{L}(d | \theta)\pi(\theta)}{\mathcal{Z}}.
\end{equation}
We use nested sampling, introduced by \citet{Skilling06} and popular for gravitational-wave data analysis due to its handling of high-dimensional spaces \citep{Veitch14}. For a thorough review of Bayesian inference in the context of gravitational-wave astrophysics, see~\citet{ThraneTalbot18}.

Our gravitational-wave transient likelihood $\mathcal{L}(d | \theta)$ is of the form
\begin{equation}
    \label{eq:likelihood}
    \mathcal{L}(d | \theta)=\frac{1}{2 \pi \sigma^{2}} \exp \left(-\frac{1}{2} \frac{(d-\mu(\theta))^{2}}{\sigma^{2}}\right),
\end{equation}
where $\mu$ is our waveform template and $\sigma$ is the detector noise amplitude spectral density\footnote{It should be noted that both $d$ and $\mu(\theta)$ are functions of frequency, and that the notation making this explicit has been omitted for brevity. There is an implied product over frequency bins.}. We assume Gaussian noise, using the noise power spectral densities $\sigma^2$ that were used to produce the GWTC-1 results. We neglect calibration uncertainty. We generate $\mu$ using \sc{SEOBNRE} \citep{SEOBNRE}, an effective one-body numerical-relativity waveform model \new{which can produce non-circular waveforms with eccentricities} in the range $0 \leq e \leq 0.2$ at \unit[10]{Hz}.
The \sc{SEOBNRE} waveform model incorporates more complex physics than the waveform used in \citet{Lower18}. It includes aligned dimensionless component spin magnitudes between -1 and 0.6, and models the merger and ringdown in addition to the inspiral. 

We carry out Bayesian inference with \sc{bilby} \citep{bilby}, using \sc{dynesty} \citep{dynesty} as our nested sampler.
We use the publicly available strain data associated with the ten binary black hole events in GWTC-1. We implement the same priors as used in GWTC-1 for almost all parameters. The exceptions are eccentricity, which is not included in GWTC-1, and aligned dimensionless component spins, which are only supported in \sc{SEOBNRE} between -1 and 0.6. Our prior on eccentricity is uniform in $\text{log}_{10}(e)$ in the range \new{$-6 \leq \text{log}_{10}(e) \leq -0.7$}. Our aligned dimensionless component spin prior is uniform between $-0.6$ and $0.6$. 
Our prior boundaries for cyclic parameters (e.g. coalescence phase, right ascension, declination), are periodic, whilst our priors for non-cyclic parameters (e.g. mass, distance) have reflective boundaries. 

Generating a posterior probability distribution demands many thousands of likelihood calculations, each requiring waveform template evaluation.
Whilst standard quasi-circular waveform models are fast enough to facilitate reasonable computation times, eccentric waveforms including merger and ringdown physics are not. \sc{SEOBNRE} takes roughly a million times longer to evaluate a GW150914-like signal than aligned-spin quasi-circular waveform model \sc{IMRPhenomD}~\citep{Khan15}. As such, \sc{SEOBNRE} is infeasible to use as our waveform template within the nested sampler calculation. Instead, we do our initial analysis with  \sc{IMRPhenomD} and \textit{reweight} our results by the eccentricity-marginalised Bayesian likelihood for \sc{SEOBNRE}, following the prescription of \cite{Payne}.

Adopting the terminology from~\cite{Payne}, our `proposal' likelihood $\mathcal{L}_{\clock}(d | \theta)$ is obtained using the quasi-circular \sc{IMRPhenomD} waveform model denoted by $\mu_{\clock}$, whilst our `target' likelihood $\mathcal{L}(d | \theta)$ is the eccentricity-marginalised likelihood calculated using the eccentric \sc{SEOBNRE} waveform denoted by $\mu$. The ratio $\mathcal{L}/\mathcal{L}_{\clock}$ provides weights, $w(d|\theta)$, that are applied to our proposal posterior samples to obtain an eccentricity-marginalised posterior distribution.

The \textit{efficiency} of reweighting is $(n_{\mathrm{effective}}/n_{\mathrm{samples}})$, where $n_{\mathrm{samples}}$ is the number of samples and 
\begin{equation}
    \label{eq:neff}
    n_{\mathrm{effective}}=\frac{\left(\sum_{i=1}^{n} w_{i}\right)^{2}}{\sum_{i=1}^{n} w_{i}^{2}},
\end{equation}
where $w_{i}$ is the weight associated with the $i^{\rm th}$ sample, is the effective number of samples after reweighting \citep{Kish}. The efficiency determines how well-sampled the posteriors are after reweighting\new{, relative to the proposal posteriors. When efficiency is low, we increase the number of proposal posterior samples to ensure a sufficient number of target samples.}

In order to obtain one-dimensional eccentricity posterior probability distributions, we construct a grid of \new{$60$} eccentricities, log-uniformly distributed between \new{$\log_{10}(e)=-6$} and $\log_{10}(e)=-0.7$ at \unit[10]{Hz to match the prior}. 
Following~\cite{Payne}, we set the time and phase by maximising the overlap between the target and proposal waveforms.
For each value of eccentricity in our grid, we compute the eccentric gravitational-wave transient likelihood using \sc{SEOBNRE}.
We take the average of this grid to find our eccentricity-marginalised likelihood.
We then draw an eccentricity at random, weighted by the cumulative likelihood grid, and add this to the unweighted posterior distribution.  
Finally, we apply our array of weights, $w$, to this eccentricity distribution to obtain the weighted posterior probability distribution for eccentricity at $\unit[10]{Hz}$.

The Bayes factor, $\mathcal{B}$, is a measure of how much one hypothesis is preferred over the other. 
It is calculated by taking the ratio of the evidences of two differing models. 
The circular evidence ${\cal L}_\circ$ is given by ${\cal L}(d|e=0)$.
The eccentric evidence is obtained by marginalising over eccentricity,
\begin{align}
    {\cal L}_e = \int de {\cal L}(d|e).
\end{align}
As such, the Bayes factor can be written
\begin{equation}
\mathcal{B} = {\cal L}_e / {\cal L}_\circ .
\end{equation}

\section{Injection study}\label{sec:injection}

\begin{table}
    \centering
    \caption{Properties of the injected signal source.}
    \begin{tabular}{|c|c|}
    \hline
         Chirp mass $\mathcal{M}$ & 28.2 $M_{\odot}$ \\
         Mass ratio $q$           & 0.86 \\
         Luminosity distance $d_L$ & $\unit[820]{Mpc}$ \\
         Eccentricity $e$         & 0.1 \\
         Dimensionless spin magnitude $\chi_1$ & 0.0 \\
         Inclination $\theta_{JN}$        & 0.4 \\
         Right ascension        & 1.02 radians\\
         Declination            & $- 0.55$ radians \\
         Phase at coalescence $\phi$ & 3.54 \\
         Polarization angle $\psi$ & 2.44 \\
         Network signal-to-noise $\rho$ & 24.9 \\
    \hline
    \end{tabular}
    \label{tab:injection_properties}
\end{table}

In order to validate our methodology, we inject a signal with eccentricity $e=0.1$ into simulated noise and recover posterior probability distributions for all parameters including eccentricity. 
We assume a two-observatory network consisting of LIGO Hanford and LIGO Livingston operating at design sensitivity, with a minimum frequency of $\unit[30]{Hz}$ to mimic the low-frequency noise from the first and second observing runs.
The parameters of the injected waveform are provided in Table \ref{tab:injection_properties}. 
The parameters are chosen to be GW150914-like, with an increased luminosity distance of $\unit[820]{Mpc}$ such that the network signal-to-noise $\rho\approx25$ to match the loudest signal-to-noise ratio in the catalogue.

\begin{figure}
    \centering
    \includegraphics[scale=0.6]{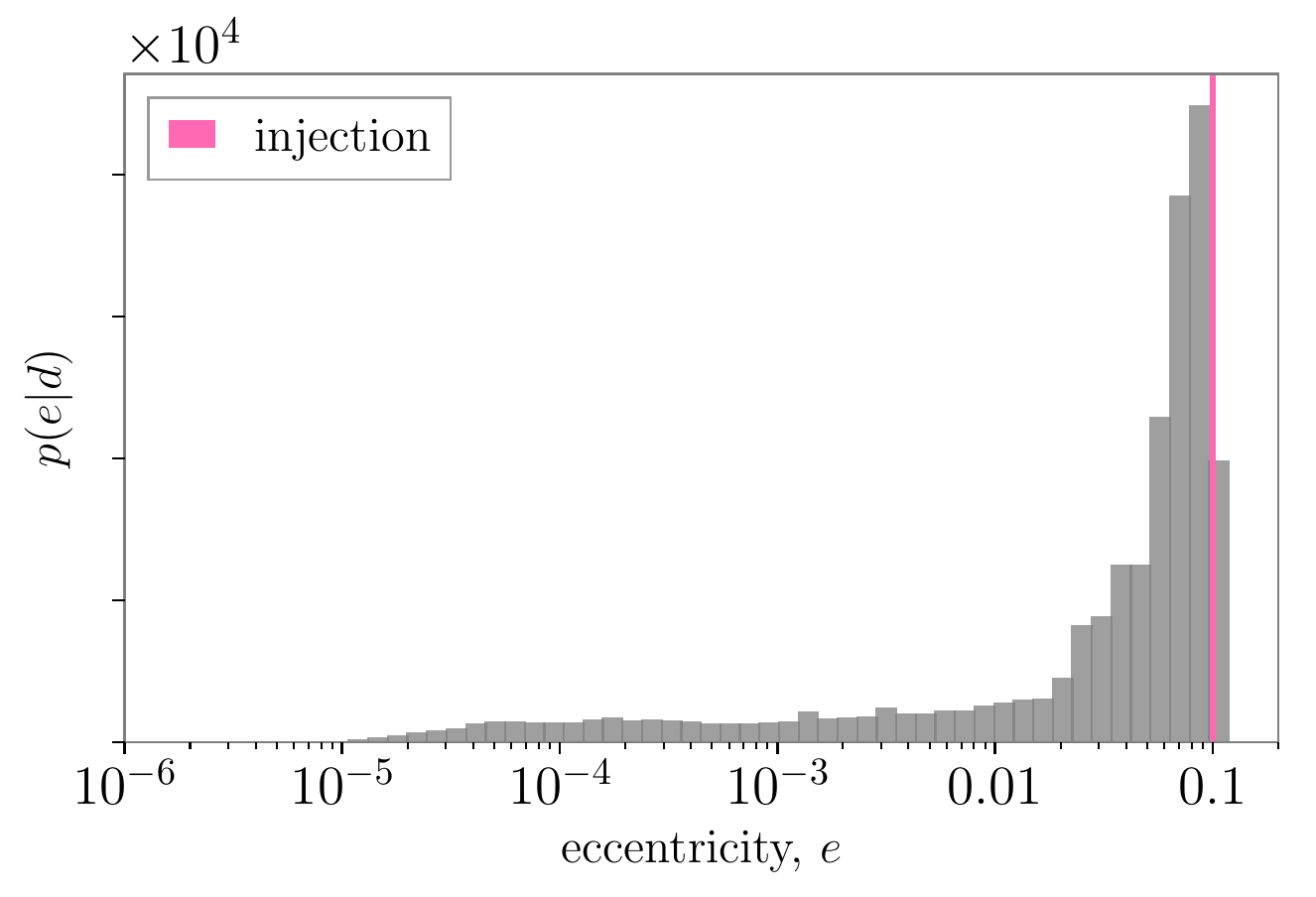}
    \caption{Reconstructed posterior probability distribution over eccentricity for our injected eccentric waveform, which has parameters as listed in Table \ref{tab:injection_properties}. For this injection, the Bayes factor for eccentricity is \new{$\text{ln}~\mathcal{B}=6.99$.}}
    \label{fig:injection_histogram}
\end{figure}

\begin{figure*}
    \label{fig:extrinsic_intrinsic}
    \centering
    \begin{subfigure}
        \centering
        \includegraphics[scale=0.35]{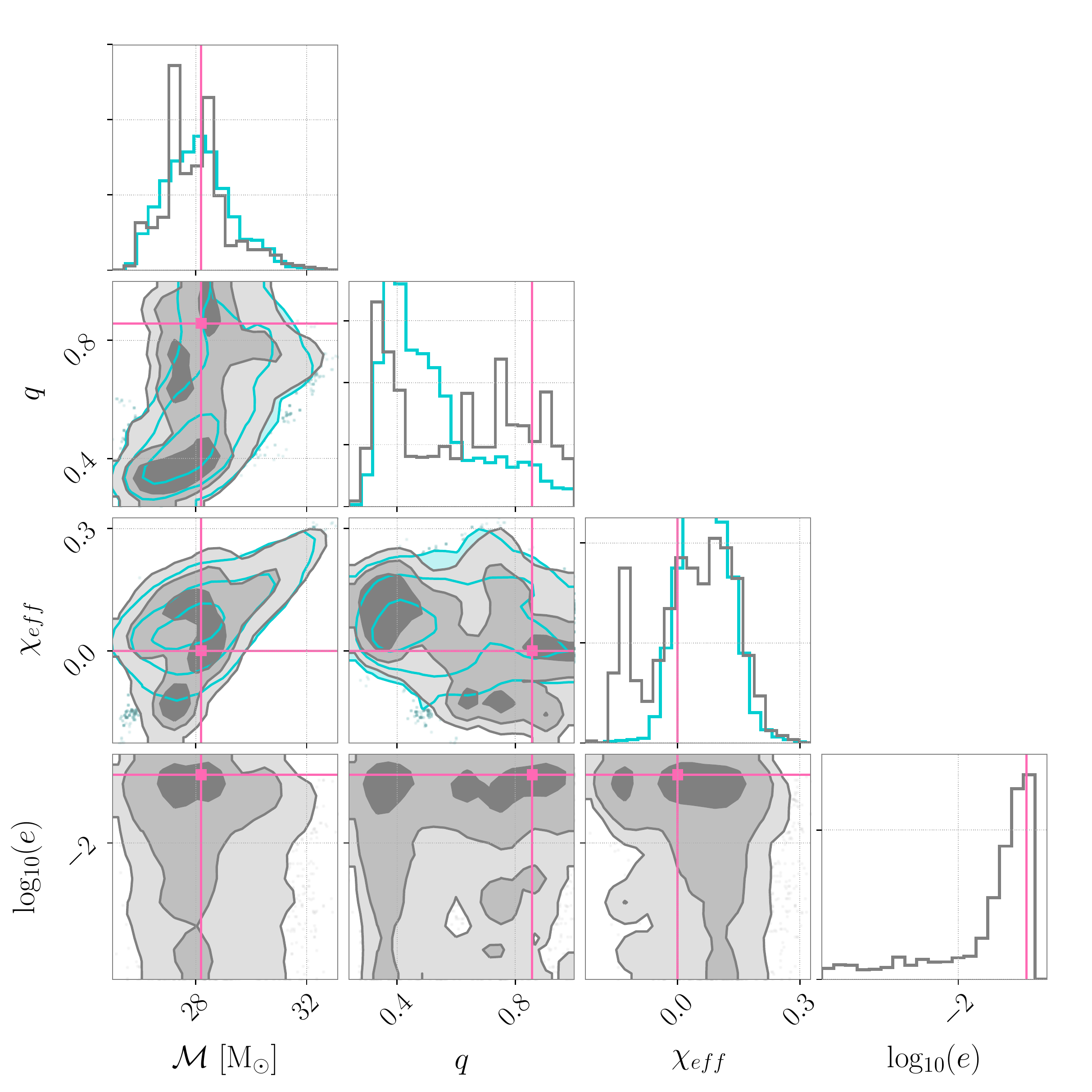}
    \end{subfigure}
    ~ 
    \begin{subfigure}
        \centering
        \includegraphics[scale=0.35]{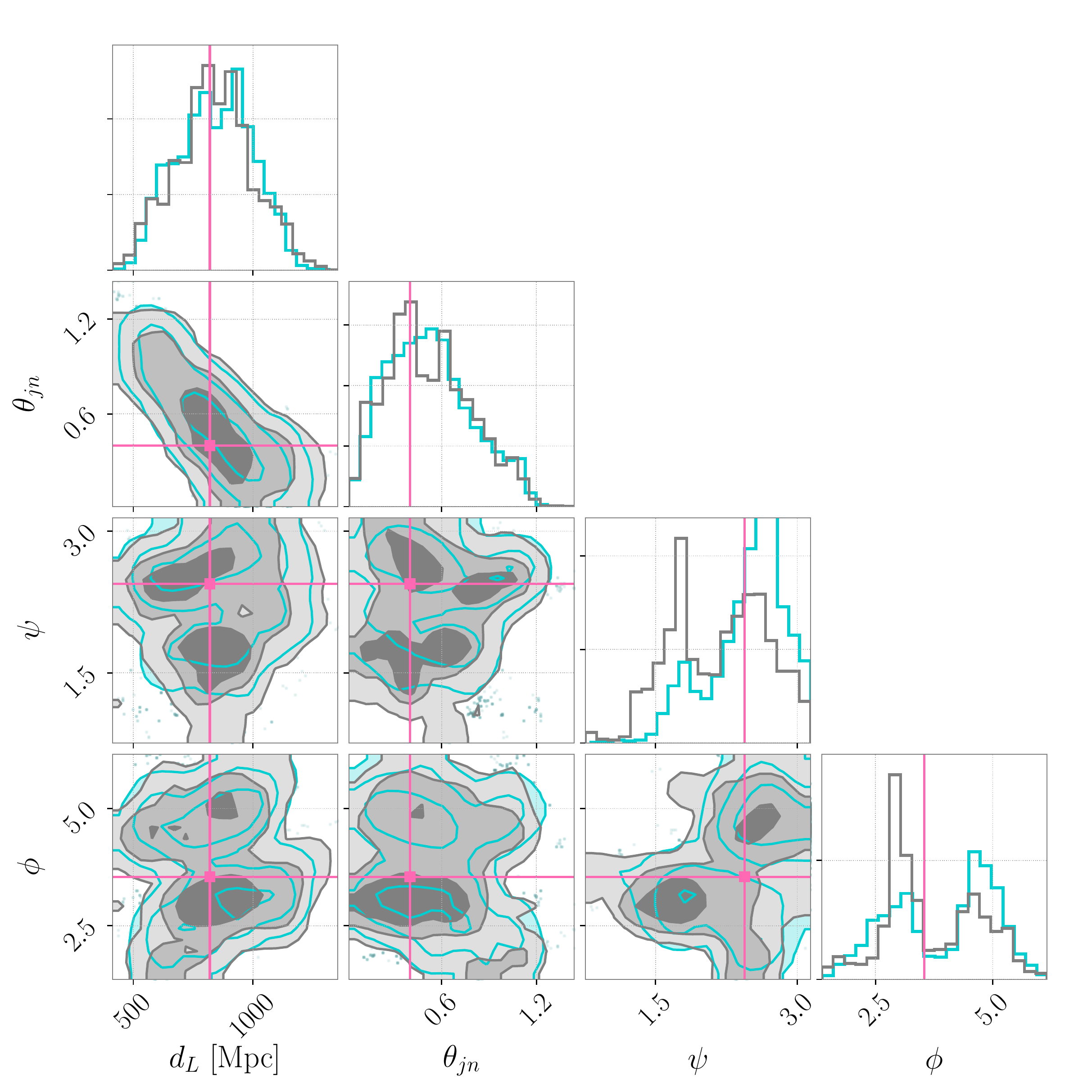}
    \end{subfigure}
    \caption{Recovered posterior probability distributions for an injected binary black hole merger signal with eccentricity $e=0.1$ and network signal-to-noise ratio $\rho\approx25$. The left-hand plot shows intrinsic parameters: chirp mass $\mathcal{M}$, mass ratio $q$, effective aligned spin $\chi_{eff}$, and log eccentricity log$_{10}(e)$. The right-hand plot contains extrinsic parameters: luminosity distance $d_L$, binary inclination angle $\theta_{jn}$, polarisation angle $\psi$, and orbital phase $\phi$. The underlying turquoise distributions are the posterior probabilities recovered using quasi-circular waveform model \sc{IMRPhenomD}~\citep{Khan15}. The grey distributions are those obtained through reweighting with eccentric waveform model \sc{SEOBNRE}~\citep{SEOBNRE}. The pink lines indicate the true parameters of the injected \sc{SEOBNRE} waveform.}
\end{figure*}

With reweighting, we obtain an eccentricity posterior that peaks at the injected value of $e = 0.1$. 
We present this posterior in Figure \ref{fig:injection_histogram}. 
As shown in Figure \ref{fig:extrinsic_intrinsic}, our initial analysis (turquoise posteriors) successfully recovers our injected signal, whilst reweighting (grey posteriors) pushes the posteriors further towards the injected values (pink lines). 
A full reweighted corner plot is available online\footnote{github.com/IsobelMarguarethe/eccentric-GWTC-1/injection}. For this injection, our reweighting has an efficiency of \new{$20\%$} and the Bayes factor for eccentricity is \new{$\text{ln}~\mathcal{B}=6.99$}.

\section{Results}\label{sec:results}

\begin{figure}
    \centering
    \hspace{-19pt}
    \includegraphics[scale=0.6]{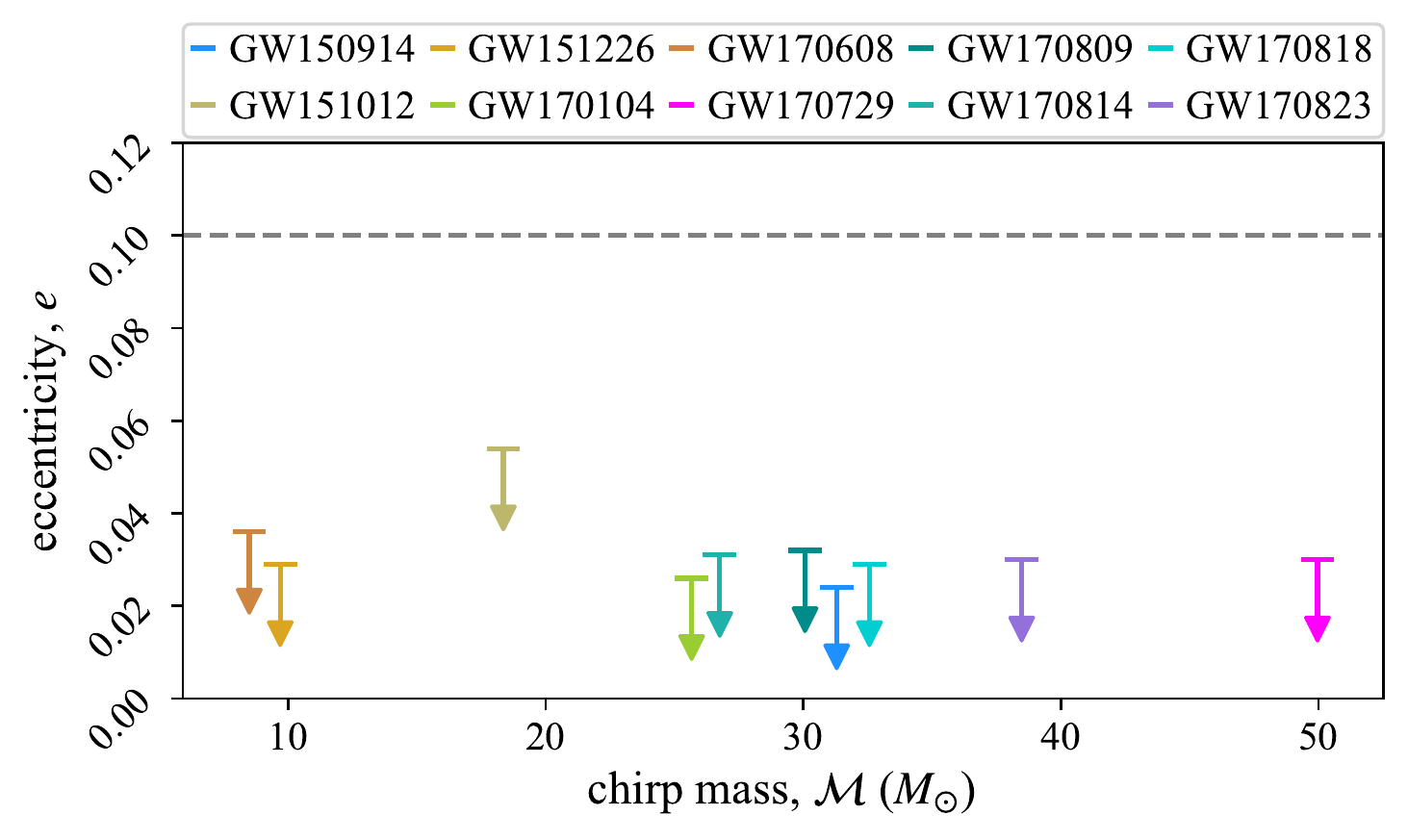}
    \caption{The 90\% credible interval upper eccentricity limit for each of the ten GWTC-1 binary black hole mergers against the mean chirp mass of the event. A dashed line is plotted at $e=0.1$, above which we should expect to see $\sim5\%$ of all mergers from globular clusters \citep{Rodriguez18b, Samsing18}.}
    \label{fig:eccentricity_limits}
\end{figure}

Our analysis yields no strong evidence for non-zero eccentricity in the first Gravitational Wave Transient Catalogue of LIGO and Virgo. We plot our 90\% confidence upper limits on eccentricity at $\unit[10]{Hz}$ against the mean of the system's chirp mass posterior in Figure~\ref{fig:eccentricity_limits}.
These limits range between \new{0.024} for GW150914 to \new{0.054} for GW151012.
\begin{table}
\centering
\caption{Upper 90\% credible interval limits on eccentricity, $e_\text{max}^{90}$, and log Bayes factors, $\text{ln}~\mathcal{B}$, for all ten binary black hole merger events published in GWTC-1 \citep{GWTC-1}. 
\label{tab:Bayes}}
\bgroup
\def\arraystretch{1.5}
\begin{tabular}{|c|c|c|}
\hline
\textbf{Event} & $e_\text{max}^{90}$ & $\text{ln}~\mathcal{B}$ \\
\hline
\textbf{GW150914} & \new{$0.024$} & \new{$-0.07$} \\
\textbf{GW151012} & \new{$0.054$} & \new{$-0.12$} \\
\textbf{GW151226} & \new{$0.029$} & \new{$-0.08$} \\
\textbf{GW170104} & \new{$0.026$} & \new{$-0.05$} \\
\textbf{GW170608} & \new{$0.036$} & \new{$-0.28$} \\
\textbf{GW170729} & \new{$0.030$} & \new{$-0.05$} \\
\textbf{GW170809} & \new{$0.032$} & \new{$-0.28$} \\
\textbf{GW170814} & \new{$0.031$} & \new{$0.05$} \\
\textbf{GW170818} & \new{$0.029$} & \new{$-0.05$} \\
\textbf{GW170823} & \new{$0.030$} & \new{$-0.12$} \\
\hline
\end{tabular}
\egroup
\end{table}
We provide event-specific log Bayes factors $\text{ln}~\mathcal{B}$ and 90\% upper confidence limits on eccentricity at $\unit[10]{Hz}$ in Table \ref{tab:Bayes}. 
Negative values of $\text{ln}~\mathcal{B}$ indicate that the data prefers the zero-eccentricity hypothesis.
Using Equation \ref{eq:neff}, we calculate the efficiency of reweighting to range from $\sim 1\%$ for GW170814 to \new{$\sim75\%$} for GW151012.
We highlight that our result does not rule out the dynamical formation channel for these mergers, since only $\sim5\%$ of globular cluster mergers are expected to be highly eccentric.
If we subsequently observe about 50 (90) events consistent with $e=0$, we can rule out the dynamical hypothesis as the only source of mergers with 90\% (99\%) confidence.

\begin{figure*}
   \centering
    \includegraphics[scale=0.8]{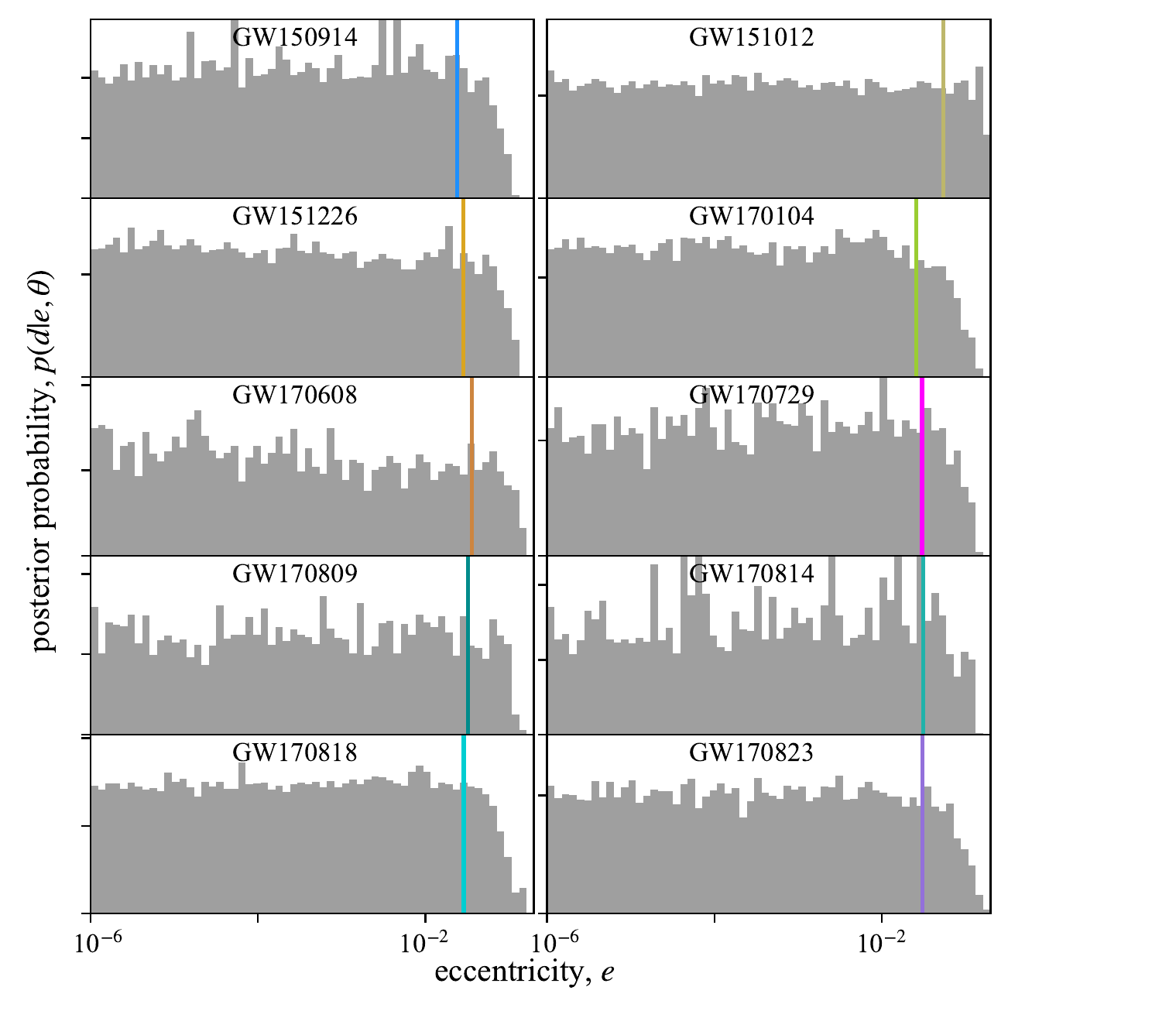}
    \caption{Eccentricity posteriors for all ten GWTC-1 events. We indicate the 90\% confidence upper limit on eccentricity at $\unit[10]{Hz}$ for each event with a vertical line, coloured to match the event-specific colours of the same limits plotted in Figure \ref{fig:eccentricity_limits}.
    \label{fig:eccentricity_posteriors}}
\end{figure*}

We present eccentricity posterior probability histograms for all ten events in Figure~\ref{fig:eccentricity_posteriors}. These are the first measurements of eccentricity in binary black holes detected with gravitational waves. 
The posteriors exhibit oscillatory behaviour because the overlap between an eccentric and a non-eccentric waveform rises and falls quasi-periodically.
Although we maximise over coalescence phase, it is not always possible to match an eccentric waveform's phase evolution to that of a quasi-circular waveform.
Thus, although the overlap follows a decreasing trend as the eccentricity increases, the overlap oscillates around this trend, and this is reflected in the likelihood.
We highlight that the heaviest event, GW170729, has the second-lowest network signal-to-noise ratio and the second-highest eccentricity upper limit. 
Gravitational waves from heavy binaries reach $\unit[10]{Hz}$ only a few cycles before the binary merges, so it is harder to distinguish a mildly eccentric signal from a quasi-circular one.
The full posteriors for all binary parameters for each event are available online\footnote{github.com/IsobelMarguarethe/eccentric-GWTC-1/events}\new{; we recover posterior probability distributions on event parameters that are consistent with those published in GWTC-1 \citep{GWTC-1}}. 

\section{Discussion}
\label{sec:discussion}
We present measurements of eccentricity for the ten binary black hole mergers in the first Gravitational Wave Transient Catalogue, finding that all of these events have eccentricities consistent with zero at $\unit[10]{Hz}$. This result does not rule out the dynamical formation channel as the primary channel for LIGO-Virgo observations.
We expect only $\sim5\%$ of globular cluster mergers to have $e \geq 0.1$ at $\unit[10]{Hz}$.
We require $\approx15$ events before it becomes more likely than not to detect eccentricity if all mergers are produced in globular clusters.
Of course, more are required if there are multiple formation channels that produce non-eccentric mergers.
Additionally, since the signal-to-noise ratio of an eccentric signal is smaller than that of its quasi-circular counterpart, quasi-circular binary signals will preferentially be detected over eccentric signals \citep{Martel99, Brown09, Randall19}. This effect will be most pronounced for $e \geq 0.1$, when the overlap between the eccentric signal and the quasi-circular signal begins to decrease rapidly.
Although unmodelled searches are able to uncover loud eccentric events (see \citet{EccentricCWB19}), current detection pipelines are likely to be preferentially detecting circular events due to as-yet-unmeasured selection effects.
Therefore, GWTC-1 may be biased towards quasi-circular binaries. 
These selection effects will be investigated in future work.

If binary neutron stars like GW170817 \citep{GW170817} efficiently form in a dynamical environment, they could retain detectable eccentricity at $\unit[10]{Hz}$ \citep{Palmese17, Andrews19}. However, studies suggest that the dynamical formation of binary neutron stars in globular clusters is highly inefficient, so mergers contributed by this channel are likely to form a small fraction of the overall binary neutron star merger rate (e.g., \citet{Grindlay05, Bae13, Zevin19ns}).
In this work, we have restricted our analysis to binary black hole systems, since modifications will need to be made to our method in order to accommodate neutron star tidal affects. 
\new{However, we intend to make our analysis applicable to binaries with neutron star components the future.}

\section{Acknowledgements}
We thank Michael Zevin, Christopher Berry and Moritz H{\"u}bner for helpful comments on the manuscript. 
This work is supported through Australian Research Council (ARC) Future Fellowships FT150100281, FT160100112, Centre of Excellence CE170100004, and Discovery Project DP180103155.
This is LIGO Document P1900228.

\bibliographystyle{mnras}
\bibliography{bib}

\end{document}